# N = 4 SUPERFIEL PHASE SPACE COORDINATES AND HAMILTONIAN QUANTIZATION

F. Assaoui[1] and T. Lhallabi[2]
*Lab/UFR High Energy Physics, Faculty of Sciences, Rabat, Morocco,
National Grouping in HEP, Focal Point, Faculty of Sciences, University Mohammed V,
B. P. 1014, Rabat, Morocco
and
The Abdus Salam International Centre for Theoretical Physics, Trieste, Italy.*

**Abstract**

The $N = 4$ superfield phase space coordinates are given in the harmonic superspace. The expressions of the $N = 4$ classical equations of motion are determined in terms of the spinorial and harmonic supercharges. Furthermore, the $N = 4$ supersymmetric actions are obtained by means of the fermionic and harmonic functionals. On the other hand, the Hamiltonian quantization is studied by performing the $N = 4$ supersymmetric action in harmonic subspace in terms of analytic $N = 4$ superfield phase space coordinates.

---

[1] Junior Associate of the Abdus Salam ICTP. fassaoui@fsr.ac.ma
[2] Regular Associate of the Abdus Salam ICTP. lhallabi@fsr.ac.ma

## I – Introduction

Local gauge invariance is a fundamental concept in field theories and there are some constraints in phase space in the gauge theories that follow the requirement of local gauge invariance. When quantizing gauge theories of the most general kind, a fundamental role is played by a rigid symmetry that transforms bosons into fermions and vice versa. This symmetry is called BRST (Becchi, Rouet, Stora, Tyutin) [1] or BFV (Batalin, Fradkin, Vilkoviski) [2, 3] and the corresponding quantization procedures are of special importance. They allow performing the quantization of general gauge and gravitational theories in a consistent way, not only with the Feynman path integral approach but also at the operator level.

However, in the generalized canonical formalism [4] all first-class constraints are independent generators of gauge transformations that generate equivalent transformations among physical states. For both the Faddeev-Senjanovic procedure [5] and BFV procedure based on BRST symmetry [6], the gauge conditions are always directly related to the first-class constraints, Furthermore, the BRST symmetry can be understood in terms of an extension of space-time to include an additional fermionic direction. This fact puts the quantization procedure for dynamical systems with first class constraints into a new geometrical framework [7]. On the other hand, N = 1 superfield formulation of the quantization program for theories with first class constraints and an exact operator formulation have been established where a phase space path integral is entirely given in terms of N = 1 superfields [8]. The BRST and canonical transformations enter on equal footing allowing to establish, on one hand, a superspace analogue of the BDF theorem [2] and on the other hand a formal derivation of the N = 1 supersymmetric lagrangian analogue of the field anti-field formalism [8]. Moreover, in the presence of second class constraints the appropriate N = 1 superfield phase space path integral provide the correct Fradkin-Senjanovic path integral measure [9] after integrating over the super partners of the ordinary fields [8]. For N = 2 superfield Hamiltonian dynamics a construction on a three dimensional superspace spanned by time t and two fermionic directions $\theta_\alpha$ is given and an N = 2 superfield phase space path integral is proposed [10]. In view of this, we find it appropriate to study the N = 4 supersymmetric extensions of the phase space coordinates, which have plenty of uses [11], by using the techniques of harmonic superspace.

The paper is organized as follows. In section II, we precise techniques of harmonic superspace and we give the N = 4 superfield phase space coordinates. The N = 4 classical equations of motion are expressed in terms of spinorial and harmonic super charges. In section III, we construct the N = 4 supersymmetric actions in terms of fermionic and harmonic functionals. In section IV, we formulate the N = 4 supersymmetric action in harmonic subspace expressed in terms of analytic N = 4 superfield phase space coordinates allowing to obtain the Hamiltonian quantization. Finally, in section V, we make our concluding remarks.

## II – N = 4 superfield phase space coordinates and harmonic superspace

The ordinary N = 4 superspace is spanned by ordinary time t and four real fermionic directions denoted by $\theta^a$, $a = 1, 2, 3, 4$; which can be represented as $\theta^{\alpha\beta}$; $\alpha, \beta = 1, 2$; due to the isomorphism group $SU(4) \sim SU(2) \times SU(2)$ [13]. The N = 4, d = 1 harmonic superspace is obtained by adding a set of harmonic coordinates parametrizing the sphere $S^2 \sim \frac{SU(2)}{U(1)}$ namely $(t, \theta^{+\alpha}, \theta^{-\alpha}, U_\beta^\pm)$ where [13]

$$\theta^{\pm\alpha} = \theta^{\alpha\beta} U_\beta^\pm \qquad (2.1)$$

with

$$U^{+\beta} U_\beta^- = 1$$
$$U_\alpha^+ U_\beta^- - U_\beta^+ U_\alpha^- = \varepsilon_{\alpha\beta} \qquad (2.2)$$

with $\varepsilon_{\alpha\beta}$ a constant antisymmetric tensor. In our formulation we kept the SU(2) indices manifest and we will not follow the formulation given in ref [12]. The spinor covariant derivatives are given by

$$\begin{aligned} D^{\pm\alpha} &= \frac{\partial}{\partial \theta_\alpha^\mp} \mp \eta^{\alpha\beta} \theta_\beta^\pm \frac{\partial}{\partial t} \\ &= \partial^{\pm\alpha} \mp \theta^{\pm\alpha} \partial_t \end{aligned} \qquad (2.3)$$

where the indices are raised with the help of the constant symmetric metric $\eta^{\alpha\beta}$ and satisfy the following relation

$$\{D^{+\alpha}, D^{-\beta}\} = 2\eta^{\alpha\beta} \partial_t \qquad (2.4)$$

Furthermore, the important ingredients of the harmonic formulism [12,13] are the harmonic derivatives which are given by

$$\begin{aligned} D^{\pm\pm} &= \partial^{\pm\pm} + \theta^\pm \theta^\pm \partial_t \\ D^0 &= \partial^0 \equiv U_\beta^+ \frac{\partial}{\partial U_\beta^+} - U_\beta^- \frac{\partial}{\partial U_\beta^-} \end{aligned} \qquad (2.5)$$

with

$$\partial^{\pm\pm} \equiv U_\beta^\pm \frac{\partial}{\partial U_\beta^\mp}$$

which satisfy

$$\begin{aligned} \left[D^{\pm\pm}, D_\alpha^\mp\right] &= D_\alpha^\pm & (1) \\ \left[D^{++}, D^{--}\right] &= D^0 & (2) \\ \left[D^0, D^{\pm\pm}\right] &= \pm 2D^{\pm\pm} & (3) \end{aligned} \qquad (2.6)$$

where the operator $D^0$ counts the $U(1)$ Cartan Weyl charge of the harmonic superfields. However, by following the N = 1 and N = 2 superfield construction [8,10], the N = 4 superfield phase space variables in harmonic superspace have the following expansion

$$q^A(t, \theta^+, \theta^-, U) = q_0^A(t, U) + \theta^{+\alpha}\psi_\alpha^{-A}(t, U) + \theta^{-\alpha}\chi_\alpha^{+A}(t, U) + \theta^{+2}U^{--A}(t, U) + \theta^{-2}V^{++}(t, U)$$
$$\theta^+\theta^- F^A(t, U) + \theta^{+2}\theta^{-\beta}\rho_\beta^{-A}(t, U) + \theta^{-2}\theta^{+\beta}\eta_\beta^{+A} + \theta^{+2}\theta^{-2}D^A(t, U) \qquad (2.7)$$

where the Grassmann parities are given by

$$\begin{aligned} \varepsilon(q^A) &= \varepsilon(q_0^A) = \varepsilon_A \\ \varepsilon(\psi_\alpha^{-A}) &= \varepsilon(\chi_\alpha^{+A}) = \varepsilon(\rho_\beta^{-A}) = \varepsilon(\eta_\beta^{+A}) = \varepsilon_A + 1 \\ \varepsilon(U^{--A}) &= \varepsilon(V^{++A}) = \varepsilon(F^A) = \varepsilon(D^A) = \varepsilon_A \end{aligned} \qquad (2.8)$$

and where the zero component $q_0^A(t, U)$ is identified with the original phase space variables. Therefore, the graded Poisson brackets for harmonic superfields are defined by [8]

$$\{f(q(t, \theta^\pm, U)), G(q(t, \theta^\pm, U))\} = f \overleftarrow{\partial}_A \omega^{AB} \overrightarrow{\partial}_B G \qquad (2.9)$$

with
$$\omega^{AB}(q(t,\theta^{\pm},U)) = \{q^A(t,\theta^{\pm},U), q^A(t,\theta^{\pm},U)\} \quad (2.10)$$
is the general symplectic superfield metric which has the following properties [8]
$$\omega^{AB} = -(-1)^{\varepsilon_A \varepsilon_B} \omega^{BA}$$
$$\omega^{AD}\partial_D \omega^{BC}(-1)^{\varepsilon_A \varepsilon_C} + \text{cyclic perm (A, B, C)} = 0 \quad (2.11)$$
$$\omega^{AB}\omega_{BC} = \delta^A_C$$
with
$$\varepsilon(\omega^{AB}) = \varepsilon_A + \varepsilon_B$$
and $\omega_{AB}$ is the inverse symplectic metric

However, the N = 4 classical equations of motion are taken to be
$$\boldsymbol{D}^{\pm\alpha}q^A(t,\theta^{\pm},U) = \{Q^{\pm\alpha}(q,t,\theta^{\pm},U), q^A(t,\theta^{\pm},U)\} \quad (2.12)$$
where $Q^{\pm\alpha}$ are spinorial and carrying $\pm 1$ $U(1)$ charges respectively. Furthermore, the existence of the harmonic derivatives (2.5) allows us to take also the other kind of equations namely
$$\boldsymbol{D}^{\pm\pm}q^A(t,\theta^{\pm},U) = \{Q^{\pm\pm}(q,t,\theta^{\pm},U), q^A(t,\theta^{\pm},U)\} \quad (2.13)$$
where $Q^{\pm\pm}$ are harmonic operators. Therefore, the application of the spinorial derivatives $\boldsymbol{D}^{\pm\beta}$ to equation (2.12) and the use of the equation (2.4) leads to
$$\partial_t q^A(t,\theta^{\pm},U) = -\{H, q^A(t,\theta^{\pm},U)\} \quad (2.14)$$
where the N = 4 superfield Hamiltonian H is defined by
$$2\eta^{\alpha\beta}H = -\boldsymbol{D}^{\{+\alpha}Q^{-\beta\}} - \{Q^{+\alpha}, Q^{-\beta}\} \quad (2.15)$$
On the other hand, by applying the harmonic derivatives to the equation (2.13) and by using (2.6.2) we obtain
$$\boldsymbol{D}^0 q^A(t,\theta^{\pm},U) = 0 = \{H_0, q^A(t,\theta^{\pm},U)\} \quad (2.16)$$
with
$$H_0 = \boldsymbol{D}^{++}Q^{--} - \boldsymbol{D}^{--}Q^{++} + \{Q^{++}, Q^{--}\} = 0 \quad (2.17)$$
Since the N = 4 superfield $q^A$ has no Cartan-Weil charge the equation (2.16) leads to the constraint (2.17). Let us now follow the procedure given in ref. [10] by multiplying (2.14) by $\theta^{\pm\alpha}$ namely
$$\theta^{\pm\alpha}\partial_t q^A(t,\theta^{\pm},U) = \{\theta^{\pm\alpha}H, q^A\} \quad (2.18)$$
and by using (2.12) we obtain
$$\partial^{\pm}q^A(t,\theta^{\pm},U) = \{\Omega^{\pm\alpha}q(t,\theta^{\pm},U), q^A(t,\theta^{\pm},U)\} \quad (2.19)$$
with
$$\Omega^{\pm\alpha}(q;t,\theta^{\pm},U) = Q^{\pm\alpha}(q;t,\theta^{\pm},U) \mp \theta^{\pm\alpha}H(q;t,\theta^{\pm},U) \quad (2.20)$$
In the same way, by multiplying (2.14) by $\theta^{\pm 2}$ and by using (2.13) we have
$$\partial^{\pm\pm}q^A(t,\theta^{\pm},U) = \{\Omega^{\pm\pm}(q,t,\theta^{\pm},U), q^A(t,\theta^{\pm},U)\} \quad (2.21)$$
where
$$\Omega^{\pm\pm}(q;t,\theta^{\pm},U) = Q^{\pm\pm}(q;t,\theta^{\pm},U) + \theta^{\pm}\theta^{\pm}H(q;t,\theta^{\pm},U) \quad (2.22)$$
As for the N = 2 supersymmetric case [10] the evolutions in $\theta^{\pm\alpha}$ is dictated by the combinations $\Omega^{\pm\alpha}$ and the evolution in time is dictated by the N = 4 superfield Hamiltonian $H$ of equation (2.15). Moreover, we note that the evolution in harmonic variables is also dictated by the combinations $\Omega^{\pm\pm}$. We precise that these harmonic quantities are note independent from $\Omega^{\pm\alpha}$.

In order to see that let us use the equations (2.13) and by applying respectively the spinorial derivatives $D^{\mp\alpha}$ we obtain

$$D^{-\alpha}D^{++}q^A = \{D^{-\alpha}Q^{++}, q^A\} + \{Q^{++}, D^{-\alpha}q^A\} \quad (2.23.1)$$

$$D^{+\alpha}D^{--}q^A = \{D^{+\alpha}Q^{--}, q^A\} + \{Q^{--}, D^{+\alpha}q^A\} \quad (2.23.2)$$

On the other hand, the application of the harmonic derivatives $D^{\mp\mp}$ respectively on the equations (2.12) leads to

$$D^{--}D^{+\alpha}q^A = \{D^{--}Q^{+\alpha}, q^A\} + \{Q^{+\alpha}, D^{--}q^A\} \quad (2.24.1)$$

$$D^{++}D^{-\alpha}q^A = \{D^{++}Q^{-\alpha}, q^A\} + \{Q^{-\alpha}, D^{++}q^A\} \quad (2.24.2)$$

by combining these equations and by using (2.6.1) we deduce that

$$Q^{+\alpha} = D^{++}Q^{-\alpha} - D^{-\alpha}Q^{++} + \{Q^{++}, Q^{-\alpha}\} \quad (2.25.1)$$

$$Q^{-\alpha} = D^{--}Q^{+\alpha} - D^{+\alpha}Q^{--} + \{Q^{--}, Q^{+\alpha}\} \quad (2.25.2)$$

Consequently the N = 4 Hamiltonian can be obtained in terms of the spinorial operators $Q^{\pm\alpha}$ or the harmonic operators $Q^{\pm\pm}$. On the other hand, the integrality conditions are given by

$$\partial_t \Omega^{\pm\alpha} + \partial^{\pm\alpha} H + \{\Omega^{\pm\alpha}, H\} = 0 \quad (2.26.1)$$

$$\partial^{\{\pm\alpha}\Omega^{\pm\beta\}} + \{\Omega^{\pm\alpha}, \Omega^{\pm\beta}\} = 0 \quad (2.26.2)$$

which are derived from (2.14) by using the fact that

$$(\partial^{\pm\alpha}\partial_t - \partial_t\partial^{\pm\alpha})q^A(t,\theta^\pm,U) = 0 \quad (2.27.1)$$

$$(\partial^{\pm\alpha}\partial^{\pm\beta} - \partial^{\pm\beta}\partial^{\pm\alpha})q^A(t,\theta^\pm,U) = 0 \quad (2.27.2)$$

and we have also

$$\partial^{\pm\pm}H + \partial_t\Omega^{\pm\pm} + \{H, \Omega^{\pm\pm}\} = 0 \quad (2.28.1)$$

$$\partial^{--}\Omega^{++} - \partial^{++}\Omega^{--} + \{\Omega^{++}, \Omega^{--}\} = 0 \quad (2.28.2)$$

which are deduced from (2.14) by using (2.13) and (2.21). The evaluation of the equation of motion (2.14) at $\theta_\alpha^\pm = 0$ leads to

$$\partial_t q_0^A(t,u) = -\{H^0, q_0^A(t,u)\} \quad (2.29)$$

with

$$H^0(q_0,t,u) = -\frac{1}{4}\eta_{\alpha\beta}\{D^{\{+\alpha}Q^{-\beta\}} + \{Q^{+\alpha}, Q^{-\beta}\}\}\Big|_{\theta^\pm=0} \quad (2.30)$$

Let us now consider the following quantity

$$X^A(t,\theta^\pm,u) = \partial_t q^A(t,\theta^\pm,u) + \{H, q^A(t,\theta^\pm,u)\} \quad (2.31)$$

and we perform a rescaling in $\theta_\alpha^\pm$ as in ref [10] namely

$$\theta_\alpha^+ \to \theta_\alpha^{+\prime} = \lambda_1 \theta_\alpha^+$$
$$\theta_\alpha^- \to \theta_\alpha^{-\prime} = \lambda_2 \theta_\alpha^- \quad (2.32)$$

Then

$$\frac{dX^A}{d\lambda_1} = \frac{d}{dt}\frac{dq^A}{d\lambda_1} + \left\{\frac{\partial H}{\partial \lambda_1}, q^A\right\} + \left\{H, \frac{\partial q^A}{\partial \lambda_1}\right\} \quad (2.33.1)$$

$$\frac{dX^A}{d\lambda_2} = \frac{d}{dt}\frac{dq^A}{d\lambda_2} + \left\{\frac{\partial H}{\partial \lambda_2}, q^A\right\} + \left\{H, \frac{\partial q^A}{\partial \lambda_2}\right\} \tag{2.33.2}$$

By using equation (2.19) we obtain

$$\frac{dq^A}{d\lambda_1} = \left\{\theta_\alpha^+ Q^{-\alpha}, q^A\right\} + \lambda_2 \left\{\theta_\alpha^+ \theta^{-\alpha} H, q^A\right\} \tag{2.34.1}$$

$$\frac{dq^A}{d\lambda_2} = \left\{\theta_\alpha^- Q^{+\alpha}, q^A\right\} - \lambda_1 \left\{\theta_\alpha^- \theta^{+\alpha} H, q^A\right\} \tag{2.34.2}$$

This leads to

$$\frac{dX^A}{d\lambda_1} = \frac{d}{dt}\frac{dq^A}{d\lambda_1} + \left\{\frac{\partial H}{\partial \lambda_1}, q^A\right\} - \left\{\theta^+ Q^-, \{q^A, H\}\right\} - \lambda_2 \left\{\theta^+ \theta^- H, \{q^A, H\}\right\} \tag{2.35.1}$$

$$\frac{dX^A}{d\lambda_2} = \frac{d}{dt}\frac{dq^A}{d\lambda_2} + \left\{\frac{\partial H}{\partial \lambda_2}, q^A\right\} - \left\{\theta^- Q^+, \{q^A, H\}\right\} - \lambda_1 \left\{\theta^- \theta^+ H, \{q^A, H\}\right\} \tag{2.35.2}$$

Furthermore, differentiating the expressions (2.34) with respect to time we derive

$$\frac{d}{dt}\frac{dq^A}{d\lambda_1} = X^B \partial_B \{\theta^+ Q^- + \lambda_2 \theta^+ \theta^- H, q^A\} + \{\theta^+ \partial_t Q^- + \lambda_2 \theta^+ \theta^- \partial_t H, q^A\}$$
$$- \{H, \{\theta^+ Q^- + \lambda_2 \theta^+ \theta^- H, q^A\}\} \tag{2.36.1}$$

and

$$\frac{d}{dt}\frac{dq^A}{d\lambda_2} = X^B \partial_B \{\theta^- Q^+ - \lambda_1 \theta^- \theta^+ H, q^A\} + \{\theta^- \partial_t Q^+ - \lambda_1 \theta^- \theta^+ \partial_t H, q^A\}$$
$$- \{H, \{\theta^- Q^+ - \lambda_1 \theta^- \theta^+ H, q^A\}\} \tag{2.36.2}$$

By combining the equations (2.36.1) and (2.36.2) with (2.35.1) and (2.35.2) respectively we obtain

$$\frac{dX^A}{d\lambda_1} = X^B \partial_B \{\theta^+ Q^- + \lambda_2 \theta^+ \theta^- H, q^A\} + \left\{\frac{\partial H}{\partial \lambda_1} + \theta^+ \partial_t Q^- + \lambda_2 \theta^+ \theta^- \partial_t H + \{\theta^+ Q^- + \lambda_2 \theta^+ \theta^- H, q^A\}, q^A\right\} \tag{2.37.1}$$

and

$$\frac{dX^A}{d\lambda_2} = X^B \partial_B \{\theta^- Q^+ - \lambda_1 \theta^- \theta^+ H, q^A\} + \left\{\frac{\partial H}{\partial \lambda_2} + \theta^- \partial_t Q^+ - \lambda_1 \theta^- \theta^+ \partial_t H + \{\theta^- Q^+ - \lambda_1 \theta^- \theta^+ H, q^A\}, q^A\right\} \tag{2.37.2}$$

Moreover, the consistency conditions (2.26.1) with the use of the rescaling (2.32) lead to

$$\frac{\partial H}{\partial \lambda_2} + \left(\theta^- \partial_t Q^+ - \lambda_1 \theta^- \theta^+ \partial_t H\right) + \left\{\theta^- Q^+ - \lambda_1 \theta^- \theta^+ H, H\right\} = 0$$

$$\frac{\partial H}{\partial \lambda_1} + \left(\theta^+ \partial_t Q^- + \lambda_2 \theta^+ \theta^- \partial_t H\right) + \left\{\theta^+ Q^- + \lambda_2 \theta^+ \theta^- H, H\right\} = 0 \tag{2.38}$$

Therefore, the equations (2.37) can be reduced to

$$\frac{\partial X^A}{\partial \lambda_1} = X^B \partial_B \left\{\theta^+ Q^- + \lambda_2 \theta^+ \theta^- H, q^A\right\}$$

$$\frac{\partial X^A}{\partial \lambda_2} = X^B \partial_B \left\{\theta^+ Q^- - \lambda_1 \theta^- \theta^+ H, q^A\right\} \tag{2.39}$$

As précised in ref [10] these homogenous equations which govern the $\lambda_1$ and $\lambda_2$ evaluation of the superfield $X^A$ imply that if $X^A(\lambda_{1,2}=0)=0$ then $X^A(\lambda_1,\lambda_2)=0$ for all $\lambda_1,\lambda_2$. Consequently if the superfield equations of motion (2.14) hold for $\theta_\alpha^\pm = 0$ they hold for all $\theta_\alpha^\pm$. In the next section we will derive suitable N = 4 supersymmetric actions which lead to the equations of motion (2.14) and the constraint (2.16).

**III – N = 4 supersymmetric actions in harmonic superspace**

By following the formalism given in ref [10] we begin this section by given the following fermionic functionals corresponding to the spinorial operators $Q^{\pm\alpha}$ namely

$$\Sigma^{\pm\alpha} = \int d\mu \left[ q^A \varpi_{AB} D^{\pm\alpha} q^B + Q^{\pm\alpha} \right] \quad (3.1)$$

and the harmonic functionals corresponding to the harmonic operators $Q^{\pm\pm}$ which are given by

$$\Sigma^{\pm\pm} = \int d\mu \left[ q^A \varpi_{AB} D^{\pm\pm} q^B + Q^{\pm\pm} \right] \quad (3.2)$$

Where $d\mu = dt\, d^2\theta^+ d^2\theta^- dU$ is the total integral measure of the one dimensional harmonic superspace and $\varpi_{AB}$ is the known expression given in ref [10] by

$$\varpi_{AB} \equiv (q^c \partial_c + 2)^{-1} \omega_{AB} = \int_0^1 \alpha\, d\alpha\, \omega_{AB}(\alpha q) \quad (3.3)$$

By using the fermionic functionals (3.1) one can define the first action $S_1$ given in terms of the functional Poisson bracket namely

$$S_1 = \frac{1}{4} \eta_{\alpha\beta} \{\Sigma^{+\alpha}, \Sigma^{-\beta}\}$$

$$= \frac{1}{4} \eta_{\alpha\beta} \Sigma^{+\alpha} \int \frac{\overleftarrow{\delta}}{\delta q^B(\mu')} d\mu'\, \Omega^{BA}(\mu',\mu)\, d\mu\, \frac{\overrightarrow{\delta}}{\delta q^A(\mu)} \Sigma^{-\beta} \quad (3.4)$$

with

$$\Omega^{BA}(\mu',\mu) = \omega^{BA}(q(t,\theta^\pm,u))\delta(t'-t)\delta^2(\theta'^+ - \theta^+)\delta^2(\theta'^- - \theta^-)\delta^2(u'-u) \quad (3.5)$$

On the other hand, one can also define the second action $S_2$ derived in terms of the following functional Poisson bracket obtained from the harmonic functionals (3.2)

$$S_2 = \frac{1}{4} \{\Sigma^{++}, \Sigma^{--}\}$$

$$= \frac{1}{4} \Sigma^{++} \int \frac{\overleftarrow{\delta}}{\delta q^B(\mu')} d\mu'\, \Omega^{BA}(\mu',\mu)\, d\mu\, \frac{\overrightarrow{\delta}}{\delta q^A(\mu)} \Sigma^{--} \quad (3.6)$$

The insertion of the definitions (3.1) and (3.2) in the expressions (3.4) and (3.6) respectively allows to obtain

$$S_1 = \int d\mu \left[ -\frac{1}{4} D^{+\alpha} q^B \eta_{\alpha\beta} \omega_{BA} D^{-\beta} q^A (-1)^{\varepsilon_A} - H \right] \quad (3.7)$$

which is written in the same way as in the N = 2 supersymmetric case [10] but in terms of N = 4 superfields and

$$S_2 = \frac{1}{4} \int d\mu \left[ D^{++} q^B \omega_{AB} D^{--} q^A (-1)^{\varepsilon_A} - H_0 \right] \quad (3.8)$$

Furthermore, the derivation of the equation of motion from $S_1$ leads to

$$\frac{\delta S_1}{\delta q^B(t,\theta^{\pm},u)} = \omega_{BA}\partial_t q^A(t,\theta^{\pm},u) - \partial_B H = 0 \qquad (3.9)$$

Which are the equations of motion of N = 1 supersymmetric case written in terms of N = 4 superfields. Concerning the second action (3.8) the derivation of the equation of motion gives

$$\frac{\delta S_2}{\delta q^B(t,\theta^{\pm},u)} = \omega_{BA} D^0 q^A(t,\theta^{\pm},u) - \partial_B H_0 = 0 \qquad (3.10)$$

which is equivalent to (2.16) since $H_0$ is nothing but the constraint (2.17). Finally, let us note the expansion of the actions (3.7) and (3.8) in terms of component fields which allows us to be left with many terms of containing auxiliary fields. In order to reduce it one has to set some constraints. This will be the subject of the next section.

**IV – N = 4 harmonic subspace and Hamiltonian quantization**

The one dimensional N = 4 harmonic superspace $\{t, \theta_\alpha^{\pm}, U_\beta^{\pm}\}$ as referred to as the central basis [13]. There exists analytic superspace which is a quotient by $\theta_\alpha^-, \alpha = 1, 2$ namely $\{t, \theta_\alpha^+, U_\beta^{\pm}\}$. In this subspace the superderivatives take the following forms

$$\begin{aligned} D_\alpha^+ &= \partial_\alpha^+ \\ D_\alpha^- &= -\partial_\alpha^- + 2\theta_\alpha^- \partial_t \end{aligned} \qquad (4.1)$$

with $\alpha = 1, 2$
So we have

$$\{D_\alpha^+, D_\beta^-\} = 2\,\eta_{\alpha\beta}\,\partial_t \qquad (4.2)$$

Then the covariant irreducibility conditions for the N = 4 superfield phase space coordinates are given by

$$D^{+\alpha} q^A(t,\theta^{\pm},u) = 0 \qquad (4.3)$$

which are recognized in the analytic basis as Grassmann analyticity conditions. This means that $q^A(t,\theta^{\pm},U)$ are independent of $\theta_\alpha^-$ namely

$$q^A = q^A(t,\theta^+,u) \qquad (4.4)$$

and have the following expansions

$$q^A(t,\theta^+,u) = q_0^A(t,u) + \theta^+ \psi^{-A}(t,u) + \theta^{+2} F^{--A}(t,u) \qquad (4.5)$$

Therefore, the N = 4 classical equations of motion (2.12) are reduced to

$$Q^{+\alpha} = 0 \qquad (4.6.1)$$

$$D^{-\alpha} q^A(t,\theta^+,u) = \{Q^{-\alpha}(q,t,\theta^+,u)\,,\, q^A(t,\theta^+,u)\} \qquad (4.6.2)$$

where the fermionic operators $Q^{-\alpha}$ are not in principle analytic. In fact, by applying the super derivatives $D^{+\beta}$ and $D^{-\alpha}$ the expressions (4.6.2) and (4.3) respectively we have

$$\begin{aligned} D^{+\beta} D^{-\alpha} q^A &= \{D^{+\beta} Q^{-\alpha},\, q^A\} - \{Q^{-\alpha},\, D^{+\beta} q^A\} \\ D^{-\alpha} D^{+\beta} q^A &= 0 \end{aligned}$$

By combining these two expressions we obtain the following equation of motion

$$2\eta^{\alpha\beta} \partial_t q^A(t,\theta^+,u) = \{D^{+\beta} Q^{-\alpha},\, q^A(t,\theta^+,u)\} \qquad (4.7)$$

where one has to set

$$-2\eta^{\alpha\beta}H = D^{+\beta}Q^{-\alpha} \tag{4.8}$$

This means that the operator $Q^{-\alpha}$ is not analytic. On the other contrary if we look for the equations (2.13) namely

$$D^{++}q^A(t,\theta^+,u) = \{Q^{++}, q^A(t,\theta^+,u)\} \tag{4.9.1}$$
$$D^{--}q^A(t,\theta^+,u) = \{Q^{--}, q^A(t,\theta^+,u)\} \tag{4.9.2}$$

and if we apply the superderivative $D_\alpha^+$ on the equation (4.9.1) then the use of (2.6.1) leads to

$$D_\alpha^+ Q^{++} = 0 \tag{4.10}$$

which means that the harmonic operator $Q^{++}$ is analytic. On the other hand, the application of the harmonic derivative $D^{++}$ and the superderivative $D_\alpha^-$ to the expression (4.6.2) and (4.9.1) respectively allows us to obtain

$$D^{++}D_\alpha^- q^A = \{D^{++}Q^{-\alpha}, q^A\} + \{Q^{-\alpha}, D^{++}q^A\}$$
$$D_\alpha^- D^{++} q^A = \{D_\alpha^- Q^{++}, q^A\} + \{Q^{++}, D_\alpha^- q^A\}$$

By combining these two expressions and by using (2.6.1) we have

$$D_\alpha^+ q^A = 0 = \{D^{++}Q^{-\alpha} - D_\alpha^- Q^{++} + \{Q^{-\alpha}, Q^{++}\}, q^A\}$$

This leads to the following constraint

$$D^{++}Q^{-\alpha} - D_\alpha^- Q^{++} + \{Q^{-\alpha}, Q^{++}\} = 0 \tag{4.11}$$

Furthermore, by applying the superderivative $D_\alpha^+$ to the expression (4.9.2) we obtain

$$D_\alpha^+ D^{--} q^A = \{D_\alpha^+ Q^{--}, q^A\}$$

and since

$$D^{--} D_\alpha^+ q^A = 0$$

the use of the expression (2.6.1) leads to

$$D_\alpha^- q^A = -\{D_\alpha^+ Q^{--}, q^A\} \tag{4.12}$$

Therefore, the identification of the expressions (4.6.2) and (4.12) implies that

$$D_\alpha^+ Q^{--} = -Q_\alpha^- \tag{4.13}$$

Finally, if we apply the derivatives $D^{--}$ and $D_\alpha^-$ on the expressions (4.6.2) and (4.9.2) respectively we obtain the following constraint

$$D^{--}Q^{-\alpha} - D_\alpha^- Q^{--} + \{Q^{--}, Q^{-\alpha}\} = 0 \tag{4.14}$$

These constraints show that the operators $Q^{\pm\pm}$ and $Q^{-\alpha}$ are not independent. In the remainder of this paper we will choose the operator $Q^{++}$ as the fundamental because it is analytic.

Now we present a classical N = 4 supersymmetric action that leads to the correct equations of motion. For that, let us first define the following covariant harmonic derivatives namely

$$\nabla^{\pm\pm} = D^{\pm\pm} - adQ^{\pm\pm} \tag{4.15}$$

by means of which the proposed equations of motion (4.9) take the compact form

$$\nabla^{++}q^A(t,\theta^+,u) = 0 \tag{4.16.1}$$
$$\nabla^{--}q^A(t,\theta^+,u) = 0 \tag{4.16.2}$$

where we have introduced the adjoint action with respect to the super Poisson bracket, $adB = \{B, .\}$. Then the N = 4 supersymmetric action in terms of the analytic operators $Q^{++}$ is given by

$$S = -\frac{1}{2}\int d\mu^{--}\left[q^A \overline{\omega}_{AB} D^{++} q^B + Q^{++}\right] + \int d\mu^{--}\left[\nabla^{++} q^A\right] h_A^B \lambda_B \qquad (4.17)$$

with $d\mu^{--} = dt\, d^2\theta^+ dU \equiv dt\, (D^-)^2 dU$ is the analytic superspace measure and where $\lambda_B(t, \theta^+, U)$ are analytic Lagrange superfield multipliers.

In the same way as the N = 2 supersymmetric case [10] we have introduced analytic superfield vielbeins $h_A^B$ corresponding to the symplectic analytic metric which is defined by

$$\omega_{AB} = (-1)^{\varepsilon_B(1+\varepsilon_D)} h_A^C \omega_{CD}^0 h_B^D \qquad (4.18)$$

where $\omega_{AB}^0$ is the analytic superfield symplectic metric in Darboux form [10]. However, in order to simplify the expansion of the action (4.17) in terms of component variables we make the following choice

$$h_A^B = \delta_A^B \qquad (4.19)$$

and we take the expansion of the analytic Lagrange superfield multiplier $\lambda_A$ as follows

$$\lambda_A(t, \theta^+, u) = \lambda_A^0 + \theta^+ \lambda_{1A}^- + \theta^+ \theta^+ \lambda_{2A}^{--} \qquad (4.20)$$

Consequently, by performing the $\theta^+$ integration in (4.17) we get

$$S = \int dtdu \left\{ q_0^A \overline{\omega}_{AB} \partial_t q_0^B + \frac{1}{4} \psi_\alpha^{-A} \omega_{AB} \eta^{\alpha\beta} \partial^{++} \psi_\beta^{-B} (-1)^{\varepsilon_B} + \eta^{\alpha\beta} D_\alpha^- D_\beta^- Q^{++} \right.$$
$$+ \left[\partial^{++} q_0^A + \{Q^{++}, q_0^A\}\right]\lambda_{2A}^{--} + \left[-\frac{1}{2}\partial^{++}\psi_\alpha^- + \psi_\alpha^{-B}\partial_B\{Q^{++}, q_0^A\}\right]\lambda_1^{-\alpha} \qquad (4.21)$$
$$\left. + \left[\partial^{++} F^{--A} - \partial_t q_0^A + F^{--B}\partial_B\{Q^{++}, q_0^A\} + \psi_\alpha^{-C}\psi_\beta^{-B}\partial_C\partial_B\{Q^{++}, q_0^A\}\eta^{\alpha\beta}\right]\lambda_A^0 \right\}$$

We note that the variation of this action with respect to the components $\lambda_{2A}^{--}$ allows to obtain the constraint (2.21) or equivalently (4.9.1) for $\theta^+ = 0$ namely

$$\partial^{++} q_0^A + \{Q^{++}, q_0^A\} = 0 \qquad (4.22)$$

On the other hand, the variation of the action (4.21) with respect to the components $\lambda_{1A}^{\alpha-}$ leads to

$$\partial^{++}\psi_\alpha^{-A} = 2\psi_\alpha^{-B}\partial_B\{Q^{++}, q_0^A\} \qquad (4.23)$$

and finally, the variation with respect to the component $\lambda_A^0$ implies that

$$\partial^{++} F^{--A} - \partial_t q_0^A + F^{--B}\partial_B\{Q^{++}, q_0^A\} + \psi_\alpha^{-C}\psi_\beta^{-B}\partial_C\partial_B\{Q^{++}, q_0^A\}\eta^{\alpha\beta} = 0 \qquad (4.24)$$

which is equivalent to the $\theta^{+2}$ term of the equations of motion (4.16.1). Therefore, inserting the above equations back into the expression (4.21) we find that at the classical level the N = 4 supersymmetric action is equivalent to

$$S_2 = \int dtdu \left\{ q_0^A \overline{\omega}_{AB} \partial_t q_0^B + \eta^{\alpha\beta} D_\alpha^- D_\beta^- Q^{++} + \frac{1}{2}\psi_\alpha^{-A}\omega_{AB}\eta^{\alpha\beta}\psi_\beta^{-C}\partial_C\{Q^{++}, q_0^B\}(-1)^{\varepsilon_B} \right\} \qquad (4.25)$$

Let us remark that in the N = 4 supersymmetric action (4.25) the spinorial partners of the phase variables remains apparent since these variables are not completely determined one in terms of the others as in the N = 2 supersymmetric case [10]. Finally at the operator level the equations of motion (4.16.1) is equivalent to the N = 4 superfield quantum equations of motion namely

$$\nabla^{++}\hat{q}^A(t, \theta^+, u) = 0 \qquad (4.26)$$

where
$$\nabla^{++} = D^{++} - (i\hbar)^{-1} ad\, \hat{Q}^{++}$$

and  (4.27)

$$ad\, \hat{Q}^{++} = \left[\hat{Q}^{++}, \ . \right]$$

Moreover, the harmonic superspace commutation relation for equal $t$, $\theta^+$ and $u$ are given by

$$\left[\hat{q}^A(t,\theta^+,u), \hat{q}^B(t,\theta^+,u)\right] = i\hbar\hat{\omega}\left(\hat{q}^B(t,\theta^+,u)\right) \quad (4.28)$$

Furthermore, one can consider a system with first class constraint by considering a Grassmann BRST generator namely

$$\Omega^{++} = \Omega^{++}(q,t,\theta^+,u) \quad (4.29)$$

and a Hamiltonian $H = H(q, t, \theta^\pm, U)$ which are taken to satisfy

$$\{\Omega^{++}, \Omega^{++}\} = 0 = \{H, \Omega^{++}\} \quad (4.30)$$

and which are combined into the operators $Q^{++}$ as

$$Q^{++}(q,t,\theta^+,u) = \Omega^{++}(q,t,\theta^+,u) - \theta^+\theta^+ H(q,t,\theta^\pm,u)$$

which is nilpotent due to the equations (4.30)

$$\{Q^{++}, Q^{++}\} = 0$$

On the other hand, the first term of the action (4.17) can be generalized to the following functional action

$$S = \int d\mu^{--} \left[K_A(q) \boldsymbol{D}^{++} q^A - Q^{++}\right] \quad (4.31)$$

where the simplistic super potential $k_A$ is related to the symplectic metric $\omega_{AB}$ as follows

$$\omega_{AB} = \left(\partial_A K_B - (-1)^{\varepsilon_A \varepsilon_B} \partial_B K_A\right)(-1)^{\varepsilon_B}$$

The study of this action and the development of the path integral quantization formalism of the N = 4 supersymmetric Hamiltonian systems with first class constraints based on an extended Poisson bracket and BRST super charges will be discussed elsewhere[14].

## *V – Conclusion*

In this paper we have introduced the techniques of harmonic superspace in one dimension in order to express the N = 4 superfield phase space coordinates. We have seen that the equations of motion can be expressed in terms of spinorial and harmonic operators $Q_\alpha^\pm$ and $Q^{\pm\pm}$ respectively. Furthermore, the N = 4 supersymmetric Hamiltonian which governs the time evolution of the N = 4 superfield phase space variables, can be derived from the combinations of the spinorial supercharges $\Omega_\alpha^\pm$ in one hand and of the harmonic supercharges $\Omega^{\pm\pm}$ in the other hand. These combinations generate translations in the $\theta_\alpha^\pm$ directions and the $u_\alpha^\pm$ harmonic variable respectively. Moreover, the N= 4 supersymmetric actions in harmonic superspace are given by introducing fermionic $\Sigma_\alpha^\pm$ and harmonic $\Sigma^{\pm\pm}$ functionals corresponding to the operators $Q_\alpha^\pm$ and $Q^{\pm\pm}$ respectively. We have seen that the expansion in terms of component field phase space coordinates contains many auxiliary terms which have to be eliminated by setting some constraints. Therefore, the N = 4 harmonic subspace in which the N = 4 superfield phase space coordinate is analytic is more useful. This allows to obtain the N = 4 supersymmetric

Hamiltonian in terms of spinorial operators $Q^{-\alpha}$ which are not analytic and which are related to the harmonic operators $Q^{--}$. These are also dependent of the harmonic operator $Q^{++}$ which is analytic. Then, the classical N = 4 supersymmetric action which leads to the correct equations of motion is given and we have seen that in the expansion of the action in terms of components, the spinorial partners of the phase space variables remains apparent. Finally, the N = 4 superfield quantum equations of motion are noted by the obvious replacements of Poisson brackets with commutators.

*Acknowledgments*

The authors would like to thank Professor G. Thompson for the interesting discussions and for reading the manuscript and the International Atomic Energy Agency and UNESCO for hospitality at the Abdus Salam International Centre for Theoretical Physics, Trieste. This work was done within the framework of the Associate and Federation Schemes of the Abdus Salam International Centre for Theoretical Physics, Trieste, Italy.

*References*

[1] C. Becchi, A. Rout and R. Stora, Comm. Math. Phys. **42** (1975) 127.
[2] E. S. Fradkin and G. A. Vilkoviski, Phys. Lett. **B 55** (1975) 244; Phys. Lett. **B 69** (1977) 309;
    E. S. Fradkin andE. S. Fradkina, Phys. Lett. **B 72** (1978) 343;
    I. A. Batalin and E.S.Fradkin, Phys.Lett . **B 122** (1983) 157.
[3] I. A Batalin and E.S.Fradkin, Ann. Inst. Henri Pioncaré **49** (1988) 145; Riv, Nuovo cim . **9** (1986) 1.
[4] Dirac PAM 1961 Lecture on Quantum Mechanics (New-York: Yeshiva University Press).
[5] P. Senjanovic, Ann. Phys. Ny **100** (1970) 227.
[6] M. Hanneau and C. Teitelboim, Quantization of Gauge System (Prenceton, NJ: Princeton University Press).
[7] I. A. Batalin, K. Bering and P. H. Damgaard, "Superield Quantization" hep-th/9708140.
[8] I. A. Batalin, K. Bering and P. H. Damgaard, Nucl. Phys. **B 515** (1998) 455; Phys.Lett . **B 446** (1999) 175.
[9] E. Deotto and E. Gozzi, Int. J. Mod. Phys. A **16** (2001) 2709;
    E. S. Fradkin, in Proc. X Winter School of Theoretical Physics Karpacs (1973) No. **207**, pp 93;
    P. Senjanovic, Annals. Phys. **100** (1976) 227 [Erratum-ibid. **209** (1991) 248].
[10] I. A. Batalin and P. H. Damgaard, '' Hamiltonian N = 2 Superfield Quantization '' hep - th / 0306153.
[11] R. de lima Rodrigues, '' The quantum mechanics system algebra: and introductory review'', hep - th / 0205017;
    C. M. Hall, arxiv: hep - th / 9910028.
[12] E. Evanov and O. Lechtenfeld, '' N = 4 sypersymmetric mechanic superspace '', hep - th / 0307111.
[13] T. Lhallabi and E. H. Saidi, Int. Jour. Mod. Phys **A3**, (1988) 187;
    A. S. Galperin, E. A. Ivanov, S. Kalitzin, V. I. Ogievetsky, E. S. Sokatchev, Class. Quant. Grav. **1** (1984) 469;
    A. S. Galperin, E. A. Ivanov, V. I. Ogievetsky, E. S. Sokatchev, Combridge University Press, (2001) 306p.
[14] F. Assaoui and T. Lhallabi, in preparation.